\documentclass[english,aps,manuscript]{revtex4}
\usepackage[T1]{fontenc}
\usepackage[latin9]{inputenc}
\usepackage{amsmath}
\usepackage{graphicx}
\usepackage{esint}

\usepackage{babel}

\begin{document}

\title{Spreading for the generalized nonlinear Schr\"{o}dinger equation with
disorder}

\author{Hagar Veksler, Yevgeny Krivolapov and Shmuel Fishman}

\address{Physics Department, Technion- Israel Institute of Technology, Haifa
3200, Israel}
\begin{abstract}
The dynamics of an initially localized wavepacket is studied for the
generalized nonlinear Schr\"{o}dinger Equation with a random potential,
where the nonlinearity term is $\left|\psi\right|^{p}\psi$ and $p$
is arbitrary. Mainly short times for which the numerical calculations
can be performed accurately are considered. Long time calculations
are presented as well. In particular the subdiffusive behavior where
the average second moment of the wavepacket is of the form $\left\langle m_{2}\right\rangle \approx t^{a}$
is computed. Contrary to former heuristic arguments, no evidence for
any critical behavior as function of $p$ is found. The properties
of $\alpha\left(t\right)$ are explored.
\end{abstract}
\maketitle
We consider the discrete nonlinear Schr\"{o}dinger Equation (NLSE) with
a random potential in one dimension:\begin{equation}
i\frac{\partial\psi_{n}}{\partial t}=-\psi_{n+1}-\psi_{n-1}+\varepsilon_{n}\psi_{n}+\beta\left|\psi_{n}\right|^{p}\psi_{n}\label{eq:1}\end{equation}
where $\varepsilon_{n}$ are i.i.d. random variables uniformly distributed
in the interval $\left[-\frac{W}{2},\frac{W}{2}\right]$, $p$ is
the degree of nonlinearity and $\beta$ is its strength. For $\beta=0$
this equation reduces to the Anderson model where all the states are
exponentially localized \cite{Lee1985}. Consequently, for $\beta=0$
if one starts with a localized wavepacket it will not spread indefinitely.
In the absence of a random potential spreading takes place for all
$p$ \cite{Sulem1999}. In fact the continuous version of \eqref{eq:1}
for $p=2$ and without the disorder is integrable \cite{Sulem1999}.
The case of $p=2$ is of experimental relevance in classical optics
\cite{Schwartz2007} and in the field of Bose-Einstein condensates,
where the NLSE is known as the Gross-Pitaevskii equation \cite{Pitaevskii2003,Dalfovo1999}.
This equation was studied extensively in the recent years, mainly
for $p=2$. In particular, the growth of the second moment was explored
and it was found (numerically) to grow subdiffusively \cite{Pikovsky2008,Flach2009,Skokos2009},
namely, for a particle initially at $n=0$,\begin{equation}
\left\langle m_{2}\left(t\right)\right\rangle =Dt^{\alpha}\end{equation}
where $m_{2}=\sum_{n}n^{2}\left|\psi_{n}\right|^{2}$ and $\alpha$
was found to be $\alpha\approx0.33$ (for $p=2$). The average $\left\langle ...\right\rangle $
denotes an average over the realizations of the random potential.

The analytical and intuitive understanding of \eqref{eq:1} is quite
poor. The simplest intuitive argument is that if a wavepacket spreads
for long enough time, the amplitude of each state becomes negligible
(since the norm, $\sum_{n}\left|\psi_{n}\right|^{2}=1$, is conserved)
and as a result the nonlinear term weakens and becomes irrelevant,
consequently, localization takes place. The difficulty with this argument
(in addition with the fact that it disagrees with numerical results
\cite{Flach2009,Pikovsky2008,Skokos2009}) is that although absolute
value of the nonlinear term becomes smaller it should be compared
to an energy scale that may decrease as well. For $p=2$ such an argument
was developed by Pikovsky and Shepelyansky \cite{Pikovsky2008} (It
is very similar to an argument that was found to work remarkably well
for another system \cite{Shepelyansky1993}). We generalize this argument
to an arbitrary value of $p$. Assuming that after some time the packet
$\psi$ is spread over $\triangle n$ states while its norm is preserved,
than typically $\left|\psi_{n}\right|^{2}\approx\frac{1}{\triangle n}$
and therefore the nonlinear term produces an energy shift of the order
$\delta E=\beta\left|\psi_{n}\right|^{p}$ that is of the order of
$\delta E\approx\triangle n^{-\frac{p}{2}}$. Comparing this term
with the typical distance between the energies of the linear problem,
$\triangle E\approx\frac{1}{\triangle n}$, gives $\frac{\delta E}{\triangle E}\approx\beta\triangle n^{-\frac{p-2}{2}}$.
Based on this argument, Pikovsky and Shepelyansky that were interested
in the case $p=2$, where $\frac{\delta E}{\triangle E}\approx\beta$
concluded that there is a critical value denoted by $\beta_{c}$ such
that for $\beta<\beta_{c}$ the nonlinear term is negligible compared
to the level spacings of the linear problem and therefore Anderson
localization holds. For $\beta>\beta_{c}$ the levels of the linear
problem are mixed and presumably Anderson localization breaks down
and spreading takes place. From this argument it turns out that $p=2$
is a critical degree of nonlinearity and for $p>2$, $\frac{\delta E}{\triangle E}\rightarrow0$
as $\triangle n$ grows and localization holds. Existence of a critical
value of $p$ was not considered in \cite{Pikovsky2008} since only
the case $p=2$ was studied. Also for the nonlinear Schr\"{o}dinger equation
without disorder, $p=2$ has a critical meaning \cite{Barab1984,Hayashi2004}.
In the present paper \emph{no evidence for the criticallity at $p=2$
was found}. This leads one to question the validity of the arguments
implying the criticallity of $p=2$ for spreading. Recently, Flach,
Krimer and Skokos presented arguments that $\alpha=\frac{1}{p+1}$
and there is no critical value of $\beta$ or $p$ \cite{Flach2009,Skokos2009,FlachErratum}.
Their arguments are supported by some numerical calculations. It is
unclear how $\alpha$ should behave when the limit $p\rightarrow0$
is taken since in this limit localization takes place and one expects
$\alpha=0$. This is another motivation for the present work. Some
arguments presented in \cite{Pikovsky2008,Flach2009,Skokos2009} involve
assumptions on chaoticity of various modes. The present work does
not test these assumptions.

There are conjectures based on perturbation theory \cite{Fishman2009}
and rigorous results \cite{Wang2008} claiming that asymptotically
the second moment of the wave packet cannot grow faster than logarithmically
as a function of time. Nevertheless, numerical data predicts a power
law growth of the second moment. If we trust the conjectures (their
violation will be very surprising and of great interest) it is reasonable
that the available numerical data is either not asymptotic (the time
scale of this problem is unknown and therefore also the time when
the system enters the asymptotic regime) or not reliable due to computational
errors. Considering this, we concentrate on the short time behavior
of a wave packet. Our results for the long time behavior are also
presented for completeness.

In order to follow the dynamics of a wave packet, we use the SABA
algorithm, which belongs to the family of split step algorithms and
evaluates the wave packet in small steps, changing from coordinate
space to momentum space. We apply the disorder and nonlinear interaction
in the coordinates space, transform the wave to momentum space and
apply there the kinetic energy term, transform it back to the coordinate
space and so on. Nearly all numerical calculations for this problem
use such methods. Additional details on the SABA algorithm, can be
found in reference \cite{Skokos2009}. Like any numerical algorithm,
the SABA algorithm accumulates errors during the calculation which
grow with the time of the integration. We use two criteria to determine
whether our results are reliable or not: \textbf{(t1)} time reversal
and \textbf{(t2)} comparison with data which is obtained using smaller
time steps. Time reversal means integrating \eqref{eq:1} from time
0 to some later time and then integrating back to time 0. At the end
of this process (if there are no errors) we should get the initial
wave packet. To measure the accumulated errors, we define $\delta_{tr}=\sum_{n}\left|\psi_{initial}-\psi_{reversed}\right|$
and demand $\delta_{tr}<0.1$. The comparison with smaller time step
is done as follows: we calculate the second moment $m_{2}\left(t\right)$
for representative realizations and then recalculate it using smaller
time step (half of the original one). We define \begin{equation}
\delta_{m_{2}}=\frac{1}{T}\int\left|\frac{m_{2,\delta t}-m_{2,\frac{1}{2}\delta t}}{m_{2,\frac{1}{2}\delta t}}\right|dt\end{equation}
and demand $\delta_{m_{2}}<0.01$. Nearly all published numerical
calculations used a more relaxed test: \textbf{(t3)} where in \textbf{(t2)}
$m{}_{2}$ is replaced by the average over realizations.

We calculate $m_{2}$ for various values of $\beta$ and $p$ and
average over 5,000 realizations until time 1000. We verified that
\textbf{(t1)} and \textbf{(t2)} are satisfied for representative realizations.
We use time steps of 0.1, 0.02, 0.01 and 0.00025 for $\beta=0.25,$
0.5, 0.75 and 1, respectively, that were chosen to satisfy \textbf{(t1)}
and \textbf{(t2)}. In addition, data is presented for $\beta=2$ and
4 using time steps of 0.1 where \textbf{(t1)} and \textbf{(t2)} are
not satisfied. The results are shown in Fig. 1b \ref{fig 1} where
$\alpha$ is obtained from fits similar to the one presented in Fig.
1a. Only the data in the interval $500\leq t\leq1000$, that does
not involve the initial spread was used in the fit of $\alpha$. If
we choose the time interval to be $300\leq t\leq1000$ or $800\leq t\leq1000$,
our results do not change in a significant way. As we could expect,
$\alpha\left(p\rightarrow0\right)\rightarrow0$ and when $p$ is large,
$\alpha$ is very small. The maximal $\alpha$ is obtained for $p$$\approx$$\frac{1}{2}$
and nothing special happens for $p=2$. We do not see any discontinuity
for $p=0$. All the lines in Fig. 1b\ref{fig 1} have similar shape
and after forming linear transformations $\overline{\alpha}=c_{1}\alpha+c_{2}$
where $c{}_{1}$ and $c{}_{2}$ are independent of $\alpha$ and $p$
and depend only on $\beta$, all the lines approximately coincide
as shown in Fig. 2\ref{fig 2}, leading us to the conclusion that
there might be some scaling property.

\label{fig 1}%
\begin{figure}
\includegraphics[width=16cm,height=9cm]{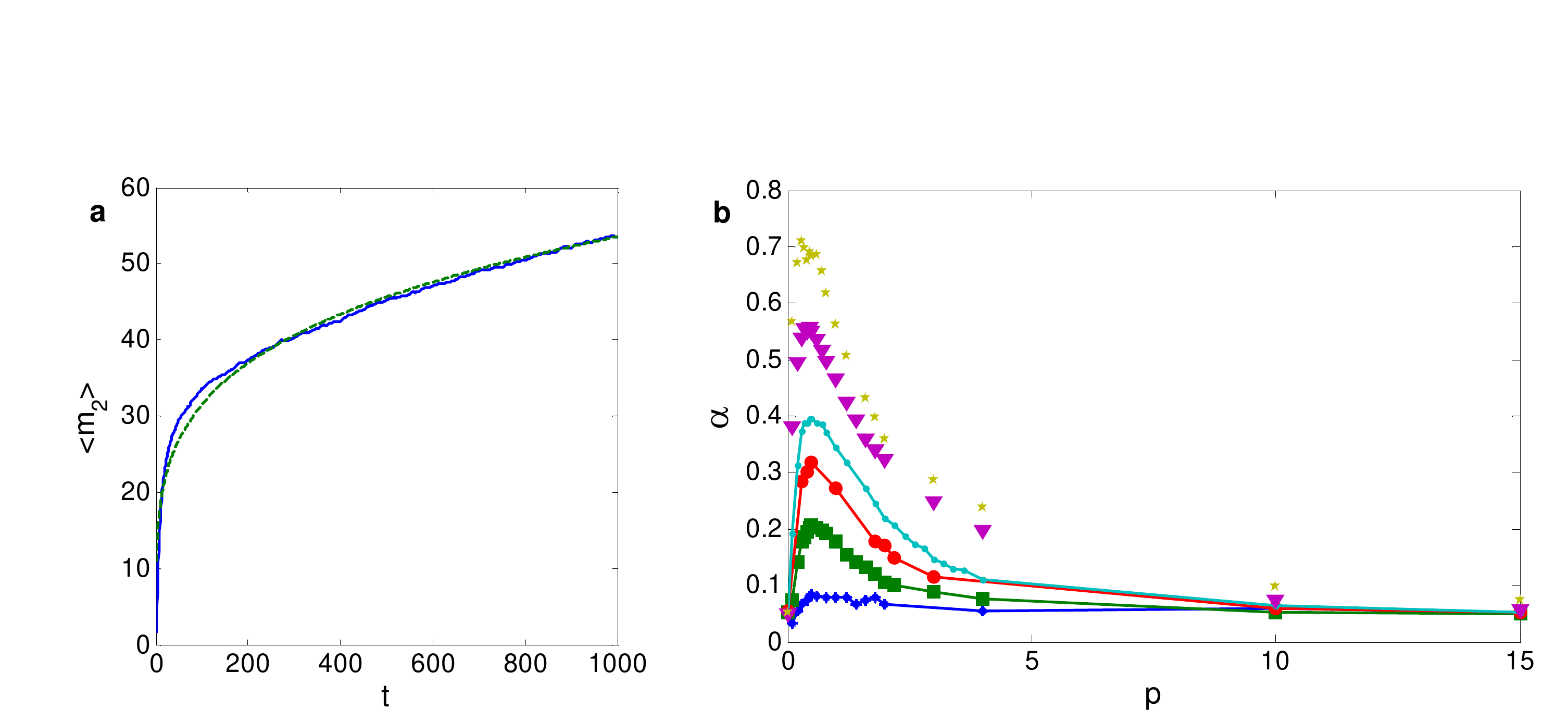}

\caption{\textbf{(a)} $\left\langle m_{2}\right\rangle $ for $\beta=1$ and
$p=2$ as a function of time. The blue solid curve is the second moment
calculated numerically and the green dashed line is the fit which
we use in order to find $\alpha$. \textbf{(b)} $\alpha\left(p\right)$
for different values of $\beta$. From top to the bottom: $\beta=4,$
2, 1, 0.75, 0.5, 0.25 (yellow stars, purple triangles, turquoise asterisks,
red circles, green squares and blue diamonds). Only the data for $\beta=0.25,$
0.5, 0.75 and $1$ where points are connected by lines satisfy \textbf{(t1)}
and\textbf{ (t2)}. For all realizations, $W=4$ and maximal localization
length is of 6 lattice sites. At the initial time the wavepacket populates
one site $\left(n=0\right)$.}

\label{Flo:prettygraph}
\end{figure}

\label{fig 2}%
\begin{figure}
\includegraphics[width=12cm,height=7cm]{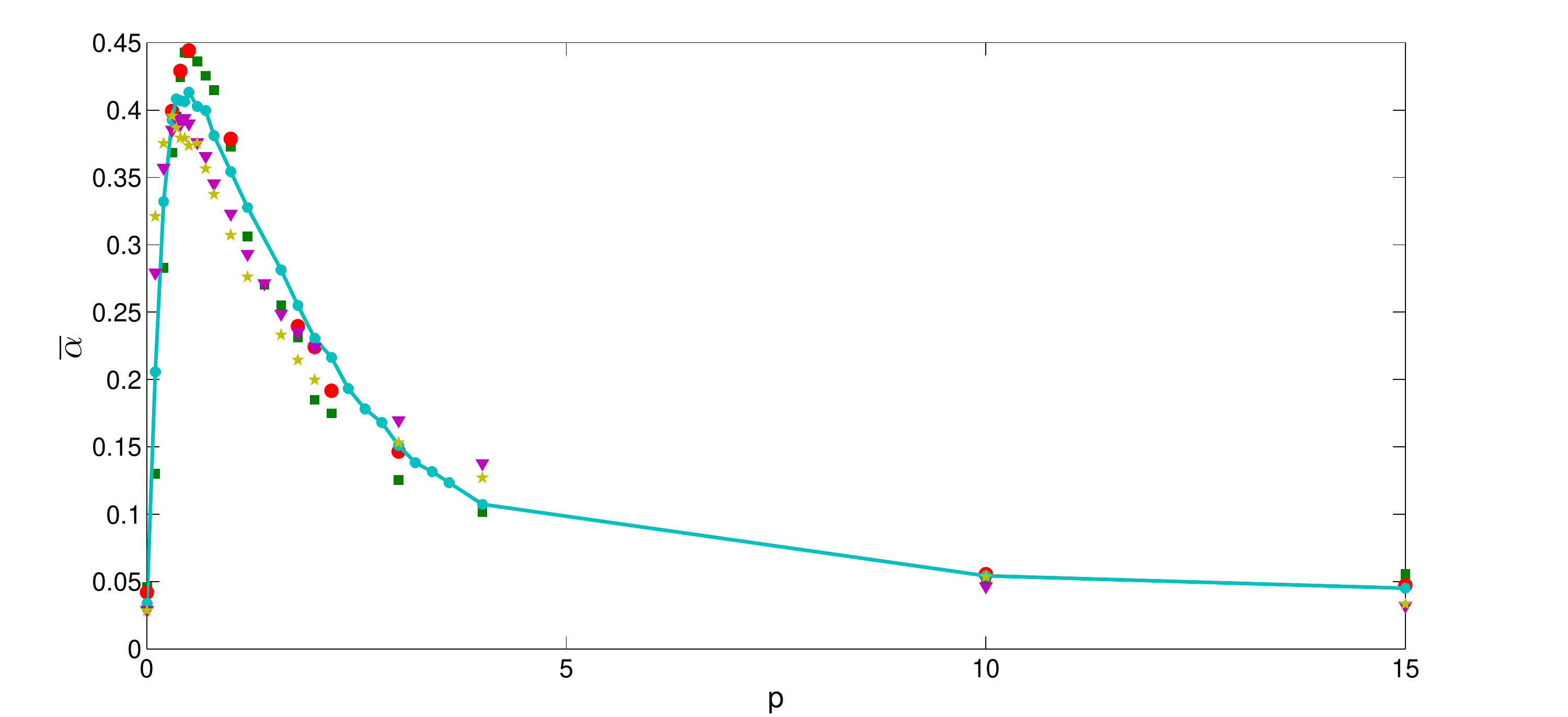}

\caption{Fig. 1b after rescaling.}

\end{figure}

In our short time runs $\left(t\leq1000\right)$ the wavepacket didn't
spread over many sites. In the case of maximal spreading ($\beta=4$,
$p=\frac{1}{2}$) the second moment reached to a maximal value of
150 and for the parameters $\beta=1$, $p=2$ the second moment was
smaller then 55, while the localization length is about 6. When we
follow the dynamics for longer times (for which \textbf{(t1)} and
\textbf{(t2)} are not satisfied but \textbf{(t3)} is satisfied) the
results support our previous conclusion that nothing critical happens
for $p=2$. We see that the wave packet spreads for all powers of
nonlinearity $p$ in a similar way, as shown in Fig. 3 for $p=0,$
1.5, 2, 2.5, 4 and $8$. Similar results are found in detailed studies
of Mulansky \cite{Mulansky2009}.

In conclusion, we have found that for short times, there is no evidence
of any critical phenomena neither for $p=0$ nor any $p=2$. This
conclusion is supported by long time calculations. In addition, we
found that $\alpha$ has a maximum for $p=\frac{1}{2}$ and there
is evidence for scaling (Fig. 2). Understanding the physics of the
$\alpha\left(p\right)$ plots, explaining why is the maximal spreading
obtained for $p=\frac{1}{2}$, explaining of the origin of the scaling
and finding the asymptotic behavior of $\alpha\left(p\right)$ are
left for future research.
\begin{acknowledgments}
We had informative discussions and communications with S. Aubry, S.
Flach, I. Guarneri, D. Krimer, M. Mulansky, A. Pikovsky, Ch. Skokos,
D.L. Shepelyansky and A. Soffer. This work was partly supported by
the Israel Science Foundation (ISF), by the USA National Science Foundation
(NSF), by the Minerva Center of Nonlinear Physics of Complex Systems,
by the Shlomo Kaplansky academic chair and by the Fund for promotion
of research at the Technion. The work was done partially while the
authors visited the Max Planck Institute in Dresden in march 2009,
and enjoined the hospitality of S. Flach.
\end{acknowledgments}
\begin{figure}
\includegraphics[width=15cm,height=7.5cm]{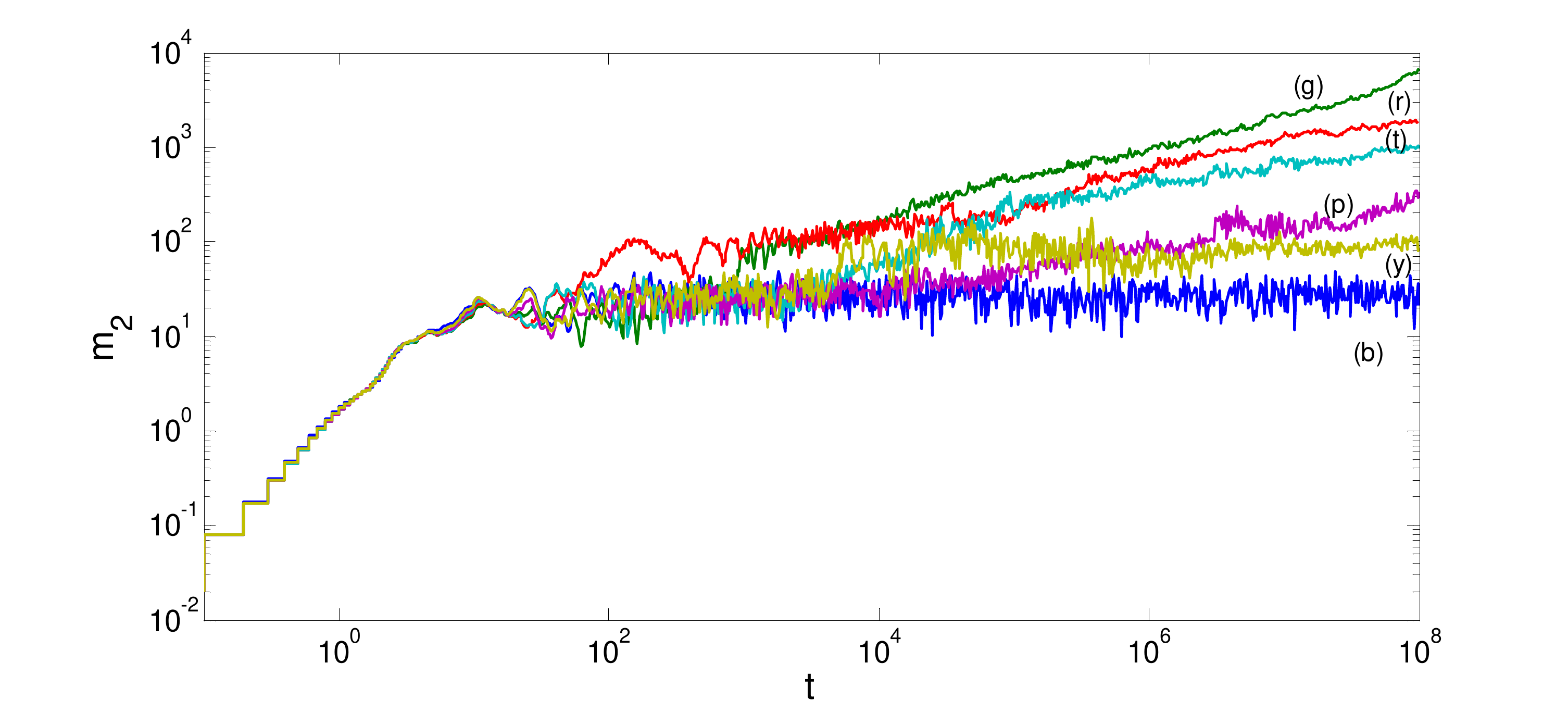}\caption{$m_{2}\left(t\right)$ for a representative realization as a function
of time for $\beta=1$ and $W=4$. From top to bottom: $p=1.5,$ 2,
2.5, 4, 8, 0 (green, red, turquoise, purple, yellow and blue).}

\end{figure}

\end{document}